\documentclass[twocolumn,showpacs,preprintnumbers,amsmath,amssymb]{revtex4-1}
\usepackage{graphicx}
\usepackage{rotating}
\begin{document}

\renewcommand{\theequation}{\thesection.\arabic{equation}}

\newcommand{\re}{\mathop{\mathrm{Re}}}

\newcommand{\lb}{\label}
\newcommand{\be}{\begin{equation}}
\newcommand{\ee}{\end{equation}}
\newcommand{\bea}{\begin{eqnarray}}
\newcommand{\eea}{\end{eqnarray}}

\title{A critical assessment of some inhomogeneous pressure Stephani models}

\author{Adam Balcerzak}
\email{abalcerz@wmf.univ.szczecin.pl}
\affiliation{\it Institute of Physics, University of Szczecin, Wielkopolska 15, 70-451 Szczecin, Poland}
\affiliation{\it Copernicus Center for Interdisciplinary Studies, S{\l }awkowska 17, 31-016 Krak\'ow, Poland}

\author{Mariusz P. D\c{a}browski}
\email{mpdabfz@wmf.univ.szczecin.pl}
\affiliation{\it Institute of Physics, University of Szczecin, Wielkopolska 15, 70-451 Szczecin, Poland}
\affiliation{\it Copernicus Center for Interdisciplinary Studies,
S{\l }awkowska 17, 31-016 Krak\'ow, Poland}

\author{Tomasz Denkiewicz}
\email{atomekd@wmf.univ.szczecin.pl}
\affiliation{\it Institute of Physics, University of Szczecin, Wielkopolska 15, 70-451 Szczecin, Poland}
\affiliation{\it Copernicus Center for Interdisciplinary Studies,
S{\l }awkowska 17, 31-016 Krak\'ow, Poland}

\author{David Polarski}
\email{David.POLARSKI@univ-montp2.fr}
\affiliation{\it Laboratoire Charles Coulomb, CNRS- Universite Montpellier II, Montpellier, France}

\author{Denis Puy}
\email{Denis.Puy@univ-montp2.fr}
\affiliation{\it LUPM, Universite Montpellier II, Montpellier, France}

\date{\today}

\input epsf

\begin{abstract}
We consider spherically symmetric inhomogeneous pressure Stephani universes, the center of
symmetrybeing our location. The main feature of these models is that comoving observers do
not follow geodesics. In particular, comoving perfect fluids have necessarily a radially
dependent pressure.
We consider a subclass of these models characterized by some inhomogeneity parameter $\beta$.
We show that also the velocity of sound, like the (effective) equation of state parameter,
of comoving perfect fluids acquire away from the origin a time and radial dependent change
proportional to $\beta$.
In order to produce a realistic universe accelerating at late times without dark energy
component one must take $\beta < 0$.
The redshift gets a modified dependence on the scale factor $a(t)$ with a
relative modification of $-9\%$ peaking at $z\sim 4$ and vanishing at the big-bang
and today on our past lightcone.
The equation of state parameter and the speed of sound of dustlike matter (corresponding
to a vanishing pressure at the center of symmetry $r=0$) behave in a similar way
and away from the center of symmetry they become negative -- a property usually encountered
for the dark energy component only.
In order to mimic the observed late-time accelerated expansion, the matter component
must significantly depart from standard dust, presumably ruling this subclass of Stephani
models out as a realistic cosmology. The only way to accept these models is to keep all
standard matter components of the universe including dark energy and take an inhomogeneity
parameter $\beta$ small enough.
\end{abstract}


\maketitle

\section{Introduction}
\label{intro}
\setcounter{equation}{0}

One of the ways to solve the dark energy problem \cite{DE} is to consider non-uniform models
of the Universe which could explain the acceleration only due to inhomogeneity
\cite{UCETolman,center}. There is a suggestion that we live in a spherically symmetric void
of density described by the Lema\^itre-Tolman-Bondi (LTB) concentric dust spheres model
\cite{LTB}. The simplest inhomogeneous cosmological models are spherically symmetric of
which category LTB models are complementary to Stephani models
\cite{stephani,dabrowski93,dabrowski95}. The former have inhomogeneous density $\varrho(t,r)$
(variable density dust shells) while the latter have inhomogeneous pressure $p(t,r)$ (variable
pressure shells).

In view of large expansion of investigations related to LTB models as nearly the only example
of an inhomogeneous cosmology, we think that it is useful to fully consider a theoretical
basis and observational validity of the complementary Stephani universes. The number of papers
about these models is tight in comparison to LTB models. In this paper we would like to fill
in this gap slightly. One of the benefits of Stephani cosmology is that it possesses a totally
spacetime inhomogeneous generalization \cite{stephani,dabrowski93} which is not the case for
LTB models and for example for Barnes models \cite{barnes73} which belong to the same
class of shear-free, irrotational, expanding (or collapsing) perfect fluid models as Stephani
models (the so-called Stephani-Barnes family). This property is of course a good step towards
developing more models of such a type - i.e. the universes which describe real inhomogeneity
of space (for a review see e.g. Ref. \cite{krasinski,BKC}) - not only those, which possess
rather unrealistic center of the Universe - something which is against the Copernican
principle. This challenge requires the proper comparison of very general inhomogeneous models
with data which was first ever done for Stephani models in Ref. \cite{dabrowski98} and for
LTB models in Ref. \cite{LTBtests}.

In general, there is a lot of inhomogeneous models which are exact solutions of the Einstein
field equations and not just the perturbations of the isotropic and homogeneous Friedmann
cosmology. Curiously, observations are practically made from just one point in the Universe
(apart from the redshift drift \cite{sandage,loeb}) and extend only onto the unique past light
cone of the observer placed on the Earth. Even the cosmic microwave background radiation is
observed from one point so that its observations prove isotropy of the Universe, but not
necessarily its homogeneity \cite{chrisroy}. As suggested in Ref.\cite{ellis} one should start
with model-independent observations of the past light cone and then make conclusions related
to geometry of the Universe, though it is difficult to really differentiate between an
inhomogeneous model of the Universe with the same number of parameters as a homogeneous dark
energy model when both fit observations.

In Ref.\cite{GSS} it has been shown that pressure inhomogeneity can mimic dark energy in the
sense that they produce the same redshift-magnitude relation. The assumption was that
inhomogeneity has dominated the universe quite recently, so it influenced only slightly the
Doppler peaks and did not influence big-bang nucleosynthesis at all. This assumption was in
agreement with the ``definition 1'' of the last scattering surface in Ref.\cite{chris}
according to which the homogeneous and isotropic radiation field on this surface was assumed
(in our prospective notation this will be equivalent to the statement that the function
$V(t,r) \to 1$ at $t \to 0$). Such a property is due to conformal flatness which does not lead
to any affect on photon paths which was also discussed recently in Ref.\cite{visser}. In the
context of inhomogeneous pressure, some generalized Ehlers-Geren-Sachs (EGS) theorems were
discussed in Ref.\cite{chris99} which said that an exactly isotropic radiation field for
every fundamental observer was possible despite there was an acceleration of the observers --
only when it vanished, the models became Friedmann. In other words, and according to this
generalized theorem, high degree of isotropy of the universe (isotropic radiation field) plus
Copernican principle did not force it to be homogeneous.

In fact, in Ref.\cite{chris} a different set of Stephani models has been studied and
discussed in the context of cosmic microwave background data. This set was actually defined by
D\c{a}browski in Ref.\cite{dabrowski93} by the formulas (43) and (44) (the scale factor and
the curvature function) as well as by the formulas (57a) and (57b) (the mass density and the
pressure). A subclass of these models was then dubbed Models I in Ref.\cite{dabrowski95} and
because of the choice of the quantity $\Delta = 0$ they offered less freedom in the choice of
parameters. A common feature of these two types of models (both with $\Delta = 0$ and $\Delta
\neq 0$) was that they had a fixed cosmic-string-like equation of state $p = -(1/3) \varrho$
at the center of symmetry. In fact, the Ref.\cite{GSS} used a version of Models II of Ref.
\cite{dabrowski95} -- the one which allowed the barotropic equation of state at the center of
symmetry (Stephani models in general do not have this property) and which are more
restrictive. Quite a large generality of the models I (with $\Delta \neq 0$) studied in
Ref.\cite{chris} eased the claim that they can fit the data despite they were significantly
inhomogeneous. In Ref.\cite{PRD13} which we will further call Paper 1, the effect of redshift
drift for the Stephani models II has been studied. The similarities and differences
between standard $\Lambda$CDM, LTB and Stephani models which can be tested by the future
astronomical data were clearly presented. In Ref.\cite{off-center14} Models I with
$\Delta =0$ as well as Models II with barotropic equation of state at the center of symmetry
were tested by Union2 supernovae data for an off-center (i.e. non-centrally placed) observer
and the maximum location of the center with respect to the observer was evaluated in each model.

In this paper, and especially in its observational part where we use the data, we restrict
ourselves to the centrally placed observers and discuss Models II with a barotropic equation
of state at the center of symmetry only. In Sec.\ref{models} we present the properties of
inhomogeneous pressure Stephani models.
In Sec.\ref{modelII} we present in details the particular model we will study, emphasizing
analogies and important differences with standard cosmological models.
In Sec.\ref{tests} we derive expressions for standard quantities used in order to constrain
our model.
Sec.\ref{data} is devoted to the comparison of our particular Stephani universe with
observational data.
Finally in Sec.\ref{conclusion} we summarize our results and give our conclusion.

\section{Inhomogeneous pressure cosmology}
\setcounter{equation}{0}
\label{models}

In Paper 1 \cite{PRD13} we presented the basic properties of the inhomogeneous pressure
Stephani universes and here we give only their most important characteristics. Mathematically,
they are the only spherically symmetric solutions of Einstein equations for a perfect-fluid
energy-momentum tensor $T^{ab} = (\varrho c^2 + p) u^a u^b + p g^{ab}$ ($\varrho$ is the mass
density, $p$ is the pressure, $g^{ab}$ is the metric tensor, $u^a$ is the 4-velocity vector,
$c$ is the velocity of light) which are conformally flat and can be embedded in a
five-dimensional flat pseudoeuclidean space \cite{stephani,dabrowski93}. A general model has
no spacetime symmetries at all, but in this paper we consider only spherically symmetric
Stephani models of which metric reads as
\be
\label{STMET}
ds^2 = - \frac{a^2}{\dot{a}^2} \left[ \frac{ \left( \frac{V}{a} \right)^{\centerdot}}
  { \left( \frac{V}{a} \right)} \right]^2 c^2 dt^2~
+ \frac{a^2}{V^2} \left(dr^2~+~r^2 d\Omega^2 \right),
\ee
where
\be
\label{VSS}
  V(t,r)  =  1 + \frac{1}{4}k(t)r^2~,
\ee
and $(\ldots)^{\centerdot}~\equiv~\partial/\partial t$. Here $a(t)$ is a generalized scale
factor, $r$ is the radial coordinate, $d\Omega$ is the metric on the sphere, and $k(t)$ is a
time-dependent curvature index which allows the Universe to ``open up'' to become negatively
curved or to ``close down'' to become positively curved.

The mass density and the pressure for a {\emph comoving} perfect fluid are given by
\begin{eqnarray}
\label{rhost}
\varrho(t) & = & \frac{3}{8 \pi G} \left[ \frac{\dot{a}^2(t)}{a^2(t)} + \frac{k(t)c^2}{a^2(t)}
\right]~,\\
\label{pst}
p(t,r) & = & \left[ -1 + \frac{1}{3} \frac{\dot{\varrho}(t)}{\varrho(t)} \frac{ \left[ \frac{V(t,r)}{a(t)}
\right]}
  { \left[ \frac{V(t,r)}{a(t)} \right]^{\centerdot}} \right] \varrho(t) c^2 \nonumber \\
  &\equiv& w_{e}(t,r) \varrho(t) c^2, \nonumber \\
  &&
\end{eqnarray}
where $G$ is the gravitational constant, and $w_{e}(t,r)$ is an effective spatially dependent barotropic
index. Of course, one can have more than one (comoving) perfect fluid as it is the case in a realistic
cosmology. We will address in great details in Section III a particular class of these models and show the
analogy and fundamental differences with standard Friedmann universes. In fact, the Stephani models admit standard big-bang singularities
($a \to 0$, $\varrho \to \infty$, $p \to \infty$), and finite density (FD) singularities of pressure \cite{sussmann,dabrowski93} which resemble sudden future singularities (SFS) \cite{barrow04,PRD05} of Friedmann cosmology. In LTB models there exist ``shell-crossing'' singularities \cite{shell}, which are of a weak type in the sense of Tipler and Kr\'olak \cite{tipler+krolak} and are similar to Friedmannian generalized sudden future singularities (GSFS) \cite{GSFS} which do not lead to geodesic incompleteness \cite{lazkoz,AIP10}.

For further discussion it is useful to mention that the components of the 4-velocity and the 4-acceleration vectors are \cite{dabrowski95}
\be
\label{velo}
  u_{t}  =  -~\frac{c}{V}~,\hspace{0.8cm} \dot{u}_{r}  =  -c~\frac{V_{,r}}{V}~,
\ee
and the acceleration scalar reads as
\begin{equation}
\dot{u}~\equiv \left( \dot{u}_{a} \dot{u}^{a} \right) ^{\frac{1}{2}}~=~
\frac{V_{,r}}{a}~.
\end{equation}
Tangent to a null geodesic vector components are \cite{dabrowski95}
\be
\label{COM}
  k^{t}  =  \frac{V^{2}}{a}~, \hspace{0.2cm}
  k^{r} =  \pm \frac{V^{2}}{a^{2}} \sqrt{1~-~\frac{h^{2}}{r^{2}}}~,\hspace{0.2cm}
  k^{\theta}  =  0~,\hspace{0.2cm}
  k^{\phi}  = h \frac{V^{2}}{a^{2}r^{2}}~,
\ee
($h$ = const.) -- the plus sign applies to a ray moving away from
the center of symmetry, the minus sign applies to a ray moving towards the center.
The constant $h$ and the angle $\phi$ between the direction of observation and the direction defined by the
observer and the center of symmetry are related by
\be
\cos{\phi} = \pm \sqrt{1~-~\frac{h^{2}}{r^{2}}}~~.
\ee
The angle $\phi$ should be taken into account when one considers off-center observers \cite{dabrowski95,off-center14}.

In Ref.\cite{dabrowski93} two classes exact spherically symmetric Stephani models were found:
\begin{itemize}
\item Models I which fulfill the condition $(V/a)^{\centerdot \centerdot}=0$
\item Models II which fulfill the condition $(k/a)^{\centerdot}=0$.
\end{itemize}
A subclass of Models I \cite{dabrowski93} which is not the one used in Ref. \cite{chris} as
well as in Refs.\cite{dabrowski95,dabrowski98}) is given by:
\be
\label{ansMI}
a(t) = \frac{1}{\gamma t + \delta}, \hspace{1.cm} k(t) = \frac{\alpha t + \sigma}{\gamma t + \delta}~~,
\ee
with the units of the constants given by: $[\alpha]$ = Mpc s$^{-1}$, $[\sigma]$ = Mpc, $[\gamma]$ = Mpc$^{-1}$ s$^{-1}$, and $[\delta]$ = Mpc$^{-1}$. The metric (\ref{STMET}) takes the form
\be
\label{metMI}
ds^2 = \frac{a^2}{V^2} \left[- \left(\frac{a}{\dot{a}}\right)^2 \left(\gamma + \frac{\alpha}{4}r^2\right)^2 c^2 dt^2 + dr^2 + r^2 d\Omega^2 \right].
\ee
Using (\ref{rhost}) and (\ref{pst}) one has for this model
\bea
\varrho(t) &=& \frac{3}{8\pi G} \left[ \frac{\gamma^2}{(\gamma t + \delta)^2} + c^2 (\alpha t + \sigma)(\gamma t + \delta) \right]~~,\\
p(t,r) &=& \frac{3c^2}{8\pi G} \left\{ \varrho(t) + \frac{1}{3} \frac{(\gamma t + \delta) + \frac{1}{4}(\alpha t + \sigma) r^2}{\gamma + \frac{\alpha}{4} r^2} \right. \nonumber \\ &\times& \left. \left[ - \frac{2\gamma^3}{(\gamma t + \delta)^3} + c^2 \alpha (\gamma t + \delta) + c^2 \gamma (\alpha t + \sigma) \right]
  \right\}~~. \nonumber
\eea
The simplest subcase of (\ref{ansMI}) is when $\sigma = \delta = 0$, since we obtain a Friedmann universe with
\be
a(t) = \frac{1}{\gamma} t^{-1}~~, \hspace{0.8cm} k(t) = \frac{\alpha}{\gamma} = const. = \alpha a(t) t~~,
\ee
and this is a phantom-dominated model with $w= -5/3$ \cite{phantom} (which has an interesting null geodesic completness features \cite{FJ}). In the limit $t \to 0$ one has a big-rip singularity with $a \to \infty$, $\varrho \to \infty$, and $p \to \infty$, while in the limit $t \to \infty$ one has $a \to 0$, $\varrho \to \infty$, and $p \to \infty$ (though it also depends on the radial coordinate $r$). If $\sigma \neq 0$ and $\delta \neq 0$, then we have the limits: a) $t \to 0$, $a \to 1/\delta$, $k \to \sigma/\delta$, $\varrho \to$ const., and
\bea
p &\to& \frac{3c^2}{8 \pi G} \left\{\frac{\gamma^2}{\delta^2} + c^2 \sigma \delta \right. \\
&+& \left. \frac{1}{3} \left[-\frac{2\gamma^3}{\delta^3} + c^2(\alpha \delta + \gamma \sigma)\right]\frac{\delta + \frac{\sigma}{4}r^2}{\gamma + \frac{\alpha}{4} r^2} \right\}~~; \nonumber
\eea
b) $t \to \infty$, $a \to 0$, $k \to \alpha/\gamma$, $\varrho \to \infty$, $p \to \infty$ and the singularities of pressure appear for $\mid r \mid = 2 \sqrt{\gamma/\alpha}$. The expansion of the curvature function $k(t)$ for small $t \to 0$ gives
\be
k(t) \approx \frac{\sigma}{\delta} - \left(\frac{\alpha}{\delta} + \frac{\sigma \gamma}{\delta^2} \right) t + O(t^2)~~.
\ee
On the other hand, for large $t \gg \delta/\gamma$ one has the expansion
\be
k(t) \approx \frac{\alpha}{\gamma} + \left( \frac{\sigma}{\gamma} -
         \frac{\alpha}{\gamma}\right) \frac{1}{t} + O(1/t^2)~~.
\ee

\section{Models II}
\label{modelII}

We will consider in detail a subclass of Models II and constrain its parameters with
observations.
For these models, the factor in front of $dt^2$ in the metric (\ref{STMET}) reduces to
$-1/V^2$, hence the line element reads \cite{dabrowski93,dabrowski95}
\be
\label{metMII}
ds^2 = - \frac{1}{V^2} dt^2  + \frac{a^2}{V^2} \left(dr^2~+~r^2 d\Omega^2 \right)
\ee
We see that this model is conformally related to a flat FLRW model.
A subclass of model II with
\be
k(t) = \beta a(t)\lb{M}
\ee
($\beta$ = constant with units $[\beta]$ = Mpc$^{-1}$) was found in Ref. \cite{stelmach01}.
We will consider models \eqref{M} here in more details and constrain them with observations.

Let us recast first the basic equations \eqref{rhost}-\eqref{pst}
in a more familiar form putting here and below $c=1$
\bea
H^2(t) &=& \frac{8 \pi G}{3}~\varrho(t) - \frac{k(t)}{a^2(t)} \nonumber \\
       &=& \frac{8 \pi G}{3}~\varrho(t) - \frac{\beta}{a(t)} \lb{H2ste} \\
\dot{\varrho}(t)~V(t,r) &=& -3H(t)~[\varrho(t) + p(t,r)]~.  \lb{drhoste}
\eea
These two equations define completely the background evolution of this cosmological
universe. Of course, we can put an arbitrary number of comoving perfect fluids, each
of them satisfying separately equation \eqref{drhoste}.

In \eqref{H2ste} and \eqref{drhoste}, we have used
\eqref{M} and we have adopted the standard notation $H\equiv \dot{a}/ a$.

It is clear from \eqref{rhost} or \eqref{H2ste} that $\varrho$ depends only on time and
has no spatial dependence. On the other hand, it is seen that the appearance of $V(r,t)$
modifies the standard energy conservation equation. This forces the pressure $p$ to depend
on the coordinate $r$. We will return to this crucial point below.

Another important point is that it is enough to find the time evolution of $\varrho$ at
the center of symmetry $r=0$ where $V(r=0,t)=1$ as this evolution does not depend on $r$.
When $V=1$ \emph{everywhere}, our model reduces to the usual FLRW model.
However, for our model \eqref{M} this is not the case, but we have $V(r=0,t)=1$. In
particular, this implies that the evolution of $\varrho(t)$ can be derived in $r=0$ from
\eqref{drhoste} using the standard conservation equation.

Similarly as Ref.\cite{stelmach01} let us assume that at the center of symmetry the standard
barotropic equation of state (EoS) $p(t) = w \varrho(t)$ holds with a time independent $w=$
const. This assumption gives
\be
\label{SJmod}
\frac{8\pi G}{3} \varrho = \frac{A^2}{a^{3(1+w)}}
\ee
where
\be
A^2 = \frac{8\pi G}{3} \varrho_0 ~a_0^{3(1+w)}~,\lb{A2}
\ee
so that \eqref{H2ste} becomes
\bea
H^2 &=& \frac{A^2}{a^{3(1+w)}} - \frac{\beta}{a} \lb{H2ste1}\\
    &=& \frac{8\pi G}{3} \left[ \varrho_0 ~\left( \frac{a_0}{a} \right)^{3(1+w)}
           + \varrho_{\beta,0} ~\frac{a_0}{a} \right]~.\lb{H2ste_w}
\eea
Here we consider a model with only one comoving perfect fluid at low redshift.
We can have more of them even at low redshift as they are certainly needed at higher
redshift in a realistic universe. In equations  \eqref{H2ste_w}
one has in mind that the comoving perfect fluid should correspond in good approximation
to dust-like matter in conventional FLRW universes.

We adopt here the conventional notation for quantities defined today.
Actually, this universe reduces completely to the standard FLRW universe at the center of
spherical symmetry $r=0$ if we identify the last term of \eqref{H2ste} or \eqref{H2ste1} for
a comoving perfect fluid with $w_{\beta}=-2/3$ (analogous to domain walls \cite{walls}) and a
trivial redefinition of its energy density $\varrho_{\beta}$
\be
\frac{8\pi G}{3} \varrho_{\beta} = - \frac{\beta}{a}~.\lb{rhobeta}
\ee
In particular, we have
\be
\beta = - \frac{8\pi G}{3} \varrho_{\beta,0}~a_0~.\lb{beta1}
\ee
The acceleration of the (generalized) scale factor $a(t)$ satisfies the equation
\bea
\frac{\ddot a}{a} &=& - \frac{4\pi G}{3}\left[ (1 + 3w)\varrho - \varrho_{\beta} \right]\\
   &=&  - \frac{4\pi G}{3}~(1 + 3w)\varrho  - \frac{\beta}{2a}~,\lb{dda}
\eea
which is trivially generalized when more comoving perfect fluids (e.g. radiation) are
taken into account. It is obvious that $\beta$ must be negative if we want \eqref{rhobeta}
to make sense. Of course one can also consider universes with positive values of $\beta$.
But in that case, this model cannot serve as an alternative to conventional dark energy
models though such models were studied \cite{nemiroff}. Note that in this analogy, in sharp
contrast to genuine comoving perfect fluids,
the equation of state parameter $w_{\beta}$ is the same \emph{everywhere} i.e. for all $r$.
We stress further that the expression for the redshift $z$ as a function of the generalized
scale factor $a(t)$ differs from the conventional one as we will see in Section \ref{tests}.
Hence the constraints on equations of state coming from luminosity-distance $d_L(z)$
get more complicated than in standard universes as we will also
see explicitly in the next Section.

However, for genuine perfect fluids, the effective equation of state parameter
$w_{e}(r,t)$ defined everywhere reads
\be
\label{weff}
w_{e}(r,t) = \left[w + \frac{\beta}{4} (1+w) a(t) r^2 \right]
\ee
with
\be
p(r,t) = w_{e}(r,t)~\varrho(t)~.\lb{prt}
\ee
Hence $w_{e}(r,t)$ is both time and space-dependent and we have in particular
$w_{e}(r=0)=w$.

Actually, the radial dependence of $w_{e}(r,t)$ is due to the radial dependence of the
fluids pressure, while at the same time the fluids energy density is homogeneous with no
spatial dependence at all. The physical reason behind this dependence is the following:
a comoving observer does \emph{not} follow a geodesic. In fact, a geodesic observer will
have a four-velocity with a non vanishing radial component, it will move in the radial
direction in addition to its movement due to the expansion.
In other words, for an observer to be comoving, one needs some extra radial force acting on
him. Analogously, for a perfect fluid to be comoving one requires some
extra radial force which is provided here by the pressure gradient due to the radial
dependence of the fluids pressure. Of course, this has implications which we will
address below.

However, let us first return to the equation of state defined at $r=0$. There is no
reason why the equation of state parameter $w$ should be constant so that we will
relax this assumption and allow for an arbitrarily time evolving equation of state
parameter $w(t)$ or $w(a)$.
Of course we still have in full generality
\be
 w(a) = w_{e}(r=0,a)~.
\ee
Due to the fact that the time evolution of $\varrho(t)$ can be found at $r=0$, the standard
result holds
\be
\varrho(a) = \varrho_0~\exp \left[ -3\int_{a_0}^a da'~\frac{1+w(a')}{a'}\right]
                           \equiv \varrho_0~f(a)~.\lb{f}
\ee
We have in particular $f(a_0)=1$.
Similarly to the Friedmann models, one can define the critical density as
$\varrho_{cr} = (3H^2)/(8\pi G)$ and the density parameter $\Omega= \varrho/\varrho_{cr}$.
We then have from \eqref{H2ste}
\be
\label{defOM}
\frac{\varrho}{\varrho_{cr}} - \frac{\beta}{a H^2} \equiv \Omega + \Omega_{\beta} = 1~,
\ee
which is valid at all times, and in particular today (at $t=t_0$), $\Omega_0 +
\Omega_{\beta,0}=1$, with (putting here explicitly speed of light $c$)
\bea
\Omega_{\beta} &\equiv& -\frac{\beta~c^2}{a~H^2}, \lb{Ombeta}\\
\beta &=& a_0 H_0^2 c^{-2} ~\left( \Omega_0 - 1 \right) < 0~.
\eea
In terms of the scale factor $a$ we have
\be
H^2(a) = H_0^2 ~\left[\Omega_0~f(a) + \Omega_{\beta,0}~\frac{a_0}{a}\right]~.
\ee
As we emphasized already several times, in a realistic cosmology one will have to introduce
at least one more perfect fluid, namely radiation. In that case the last equations above are
trivially generalized as follows
\be
\frac{\varrho}{\varrho_{cr}} +  \frac{\varrho_{\rm rad}}{\varrho_{cr}} - \frac{\beta ~c^2}{a H^2}
                              \equiv \Omega + \Omega_{\rm rad} + \Omega_{\beta} = 1~,
\ee
and in particular today $\Omega_0 + \Omega_{\rm rad,0} + \Omega_{\beta,0}=1$,
\be
H^2(a) = H_0^2 ~\left[\Omega_0~f(a) + \Omega_{\rm rad,0}~\left( \frac{a_0}{a} \right)^4
                                        + \Omega_{\beta,0}~\frac{a_0}{a}\right]~,\lb{Hfull}
\ee
and finally
\be
\beta = a_0 H_0^2 c^{-2}~\left( \Omega_0 + \Omega_{\rm rad,0} - 1 \right) < 0~.
\ee
One may wonder why we do not append any suffix to the first term, like we do with the radiation
term. This is because the first term will not behave like dust-like matter, not even at $r=0$
while the radiation component does (by choice) at $r=0$ with
\bea
w_{e,{\rm rad}}(r,t) &=& \left[w_{\rm rad} + \frac{\beta}{4} (1+w_{\rm rad}) a r^2 \right]\\
&=& \frac{1}{3} ~\left[ 1 + \beta a r^2 \right]~.
\eea
and
\be
p_{\rm rad}(r,t) = w_{e,{\rm rad}}(r,t)~\varrho_{\rm rad}(t)~.\lb{prtrad}
\ee
Like for any comoving perfect fluid, $w_{e,{\rm rad}}(r,t)$ is both time and space-dependent
and we have in particular
\be
w_{e,{\rm rad}}(r=0)=w_{\rm rad}=\frac{1}{3}~.
\ee
The standard behaviour for radiation holds at $r=0$.

As we have mentioned above, comoving observers do not follow geodesics, the four-velocity
of geodesic observers will have a non-vanishing radial component. For this reason, the
three-momentum $|\bf{\vec{p}}|$ of a free particle will not evolve like $\propto V/a$.
This can have important consequences for the thermal history of our universe, e.g. the
distribution function of relics. Clearly all these effects should remain rather small for
an acceptable cosmology.

There is another very interesting point. For perfect fluids with a barotropic equation
of state of the type $p(r,t) = w_e(r,t)~\varrho(t)$,
it is straightforward to compute the corresponding velocity of sound $c_S$.
Specializing to our model with $k(t)=\beta~a(t)$, the following result is obtained
\bea
&&c_S^2(r,a) = w + \frac{\beta}{4} (1+w) a r^2 - \nonumber \\
   &-& \frac{a}{3(1+w)}\left[ \frac{dw}{da}
        + \frac{d}{da}\left( \frac{\beta}{4} (1+w) a r^2 \right) \right]~,\lb{cS}
\eea
where $w$ is the equation of state parameter at $r=0$. This expression simplifies
when $w$ has no time dependence, viz.
\bea
&&c_S^2(r,a) = w + \left( w + \frac{2}{3} \right) \frac{\beta}{4} a r^2 \nonumber \\
&=& c_S^2(r=0) + \left( c_S^2(r=0) + \frac{2}{3} \right)\frac{\beta}{4} a r^2~.\lb{cSconst}
\eea
For a perfect fluid behaving like radiation today at $r=0$, we obtain
\be
c_{S,{\rm rad}}^2(r,a) = \frac{1}{3} + \frac{\beta}{4} a r^2~,\lb{cSrad}
\ee
and for a perfect fluid behaving like dustlike matter today at $r=0$, we get
\be
c_{S,{\rm m}}^2(r,a) =  \frac{\beta}{6} a r^2~.\lb{cSm}
\ee
We see firstly that the velocity of sound, even for constant $w$, develops both radial and time
dependence. Secondly, for dust ($w(r=0)=0$), if the parameter $\beta$ is negative, not only the
pressure but also the velocity of sound squared will become negative for $ar^2\ne 0$. Such a
situation is encountered already in standard cosmology for a dark energy component with constant
negative $w$. But here we face this conceptual problem even for dust.
The departure from the standard velocity of sound resulting from \eqref{cSm} must be addressed
when considering the formation of structure. Of course these problems could be addressed in the
same way as for dark energy clustering.

The departure coming from \eqref{cSrad} and \eqref{cSm} could further affect the CMB sound
horizon and acoustic oscillations but we will show that this effect is extremely small.
Clearly all these effects may be acceptable for $\beta$ sufficiently small.

Finally, it is interesting to note that very generally for any perfect fluid with constant
$w$, the same standard velocity of sound is obtained both at $r=0$ and at the time of the
Big Bang $a=0$.

It is seen from \eqref{cSconst} that the departure $\Delta c_S^2$ from the
standard sound velocity for a barotropic perfect fluid with constant $w$ is
proportional to (putting explicitly $c$ again)
\bea
\Delta c_S^2(r,a) &\propto& \beta a r^2 c^2 \nonumber \\
   &\propto& - \Omega_{\beta,0}\frac{a}{a_0} H_0^2 (a_0 r)^2~.\lb{Delcs}
\eea
As expected, the quantity $H_0 a_0 r$ has unit of velocity and can be
conveniently estimated from
\be
H_0 a_0 r = 100 h \frac{a_0 r}{{\rm Mpc}} {\rm km/s}~,\lb{Delcs1}
\ee
with $h\equiv H_0/(100 {\rm km/s/Mpc})$.
A rough estimate of \eqref{Delcs} using \eqref{Delcs1} indicates that \eqref{Delcs}
does not become too large in observational data.

\begin{figure}
\resizebox{80mm}{!}{\includegraphics{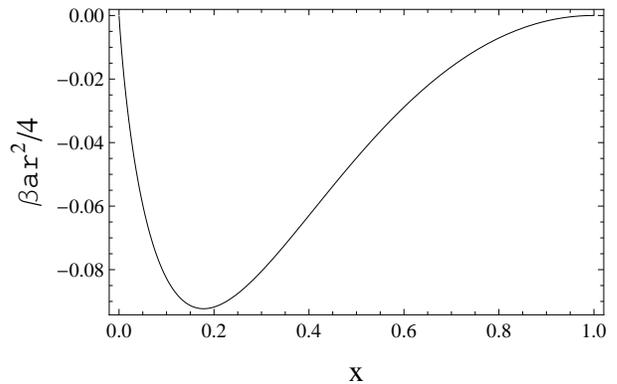}}
\caption{The quantity $\frac{1}{4}\beta a r^2$ computed on our past lightcone is shown as a
function of $x \equiv a/a_0$. Up to a constant of order one it gives the change in
the equation of state parameter and in the speed of sound for dust ($w(r=0)=0$) and
radiation ($w(r=0)=\frac{1}{3}$). It also gives the relative change in the redshift as a
function of $x$.
It vanishes both at the Big Bang and today and and it is seen
to have an extremum of about $9\%$ at $z\sim 4$. Here it is plotted
for the particular set of the cosmological parameters $\Omega_{\beta,0}=0.68$ and
$w\equiv w(r=0)=-0.08$. At very small values of $x\lesssim 10^{-3}$,
we have $\frac{1}{4} |\beta| a r^2\lesssim 10^{-3}$ }
\label{fig1}
\end{figure}

Actually, this quantity can be accurately computed on our past lightcone for given
cosmological parameters. Indeed, extending the results of \cite{stelmach01,GSS,dabrowski93}
when we have a component with a time dependent equation of state parameter $w(r=0)=w(a)$
and taking further into account a radiation component, we have for a lightray reaching
us ($r=0$) today
\be
\label{rx}
r(x)  = \frac{c}{H_0 a_0} I(x)
\ee
where
\be
x \equiv a/a_0~,~I(x) \equiv \int_x^1 \frac{dx'}{\sqrt{\Omega_0 f(x') x'^4 +
                           \Omega_{\rm rad,0} + \Omega_{\beta,0} x'^3}}~.\lb{Ix}
\ee
The quantity $\frac{1}{4}\beta a r^2$ computed on our past lightcone is shown as a function of
$x \equiv a/a_0$ on figure \ref{fig1} and we see that it has a minimal value of about $9\%$
at $z\sim 4$.
As expected  it vanishes both at the Big Bang and today. It is even very
small in the primordial era of the universe as well as at late times.

A few words about the redshift in such universes will be added
in details in the next Section. We will show that it is a modified function of $x$
(see \eqref{redshift}), viz.
\bea
1 + z &=& x^{-1} \left( 1 + \frac{1}{4} \beta a r^2(x) \right)\\
      &=& x^{-1} - \frac{\Omega_{\beta,0}}{4}~I^2(x)~,  \lb{zste}
\eea
hence
\bea
\frac{1+z(x) - \frac{a_0}{a} }{ \frac{a_0}{a} } &=& \frac{1}{4} \beta ~a ~r^2(x)\\
               &=& - \frac{\Omega_{\beta,0}}{4}~x~I^2(x)\lb{bar2}
\eea
So we have an elegant result that the quantity $\frac{1}{4} \beta ~a ~r^2$ (times $c^2$)
gives the order of magnitude of the change on our past lightcone in the velocity of sound,
in the equation of state parameter of comoving perfect fluids, in the modification  of the
metric (through the function $V$) compared to a flat Robertson-Walker metric, and finally the
relative change of the redshift $z(x)$ as a function of $x$.
We can conclude from the discussion above that the differences in the standard
dependence of the redshift on $x$ remains rather small, though not negligible
(see figure \ref{fig1}).

It is quite clear from the results of this section that this inhomogeneous universe
cannot serve as an alternative to dark energy model. Actually, the crucial problem
is that $\beta$ dependent term behaves like a perfect fluid with an equation of
state parameter equal to $-\frac{2}{3}$.
This last property is quite a general property of Models II as defined in
Ref.\cite{dabrowski93} which comes from the condition $(k/a)^{\centerdot}=0$. It is worth
emphasizing that Models I of this reference used in Ref. \cite{chris} in general do not have
such a property (cf. their Eq. (6) which allows this property only if $\Delta =0$). Instead,
they have in general a non-barotropic equation of state which at the center of symmetry
reduces to a barotropic one which describes network of cosmic strings with an equation of
state parameter equal to $-\frac{1}{3}$ \cite{dabrowski93}. In order to comply with the data,
the ``matter'' component will be forced to behave very differently from standard dust already
at the background level. It is nevertheless interesting to study how close to a viable
universe this universe can be.
In that case we must obviously have $\beta<0$ and $\Omega_{\beta,0} \sim 1$.
The effects discussed in this section are small enough at very high redshifts so that CMB
cosmological constraints can be ``translated'' in good approximation to our model in a
self-consistent approach. Doing this analysis will give us an opportunity to derive the
expressions for the redshift, the luminosity, and the angular diameter distance in the
Stephani models under study.
We emphasize that it is the radial dependence of $g_{00}$, specific to Stephani models,
which affects sound velocities, the redshift as well as cosmic distances.


\section{Observational constraints}
\setcounter{equation}{0}
\label{tests}

We should first consider the redshift -- a crucial theoretical and observational quantity.
We proceed as in Friedmann cosmology, and consider an observer located at $r=r_0=0$ at
coordinate time $t=t_0$. The observer receives a light ray emitted by a comoving source
at $r=r_e$ at coordinate time $t=t_e$ and the redshift reads
as \cite{KS,dabrowski95}
\be
\label{redshift}
1+z = \frac{(u_a k^a)_e}{(u_a k^a)_o} = \frac{\frac{V(t_e,r_e)}{a(t_e)}}{\frac{V(t_0,r_0)}{a(t_0)}}
            = \frac{a(t_0)}{a(t_e)} V(t_e,r_e)\equiv \frac{a_0}{a_e} V_e,
\ee
where we have used
\begin{eqnarray}
\label{rs}
u_a k^a = -\frac{1+\frac{1}{4}k(t)r^2}{a(t)} = -\frac{V}{a}~.
\end{eqnarray}
obtained from (\ref{velo}) and (\ref{COM}). At this stage we emphasize the following very
important point. When observational data are given
in terms of redshift, what is meant by redshift is the ratio between the observed
wavelength $\lambda_0$ at time $t_0$ (at $r=0$) and the wavelength $\lambda_e$ at
emission time $t_e$ for light emitted by a comoving source, i.e.
$\lambda_0/\lambda_e = 1 + z$.

If we want to use the observational data, we have to make sure that the redshift defined
in \eqref{redshift} retains this physical meaning.
While in standard cosmology we have $\lambda_0/\lambda_e = a_0/a_e$, in
our model we have
\be
\frac{\lambda_0}{\lambda_e} = \frac{a_0}{a_e} V_e~,
\ee
which indeed corresponds to the expression for $1+z$ defined in \eqref{redshift}.
We stress once more that this is true for light emitted by \emph{comoving} sources. And this
brings us to the following interesting question. While the comoving fluid is comoving due to a
radial dependent pressure, if matter clusters in the course of expansion it is not clear that
clustered objects remain comoving. As we have mentioned in the Section \ref{modelII}, a test particle
following a geodesic will not be comoving. This is clearly a very hard problem to solve and is
beyond the scope of this work. We can only assume that the departure from a comoving movement
is small enough that we can use compact objects like SNIa as comoving objects and check that
this is self-consistent with the obtained best-fit models.

\begin{center}
a) luminosity distance
\end{center}

The luminosity distance versus redshift relation reads \cite{stelmach01,GSS}
\be
\label{dL}
d_L = (1+z)a_0 r~~,
\ee
and the distance modulus is
\be
\mu(z)=5\log_{10}d_L(z)+25.
\ee
Interestingly, the angular diameter distance $d_A$ and the luninosity distance $d_L$
are related to each other in our Stephani universe (\ref{metMII}) exactly like in Friedmann universe, viz.
\be
d_A = \frac{a_e}{V_e} r_e = (1 + z)^{-2} d_L~.\lb{dA}
\ee
Hence, the relation between both distances does not allow to discriminate between
our model and the standard FLRW model.

Using the definition of redshift (\ref{redshift}), and using \eqref{rx}, one can write the redshift
along null geodesics as the function of $x$ \cite{GSS} as
\be
z(x) = \frac{1}{x} - 1 - \frac{\Omega_{\beta,0}}{4} \left[ \int_x^1 \frac{dx'}{\sqrt{\Omega_0 f(x') x'^4 +
                           \Omega_{\rm rad,0} + \Omega_{\beta,0} x'^3}} \right]^2~~.\lb{zx}
\ee
Inverting this function numerically gives $x(z)$. Hence, the luminosity distance (\ref{dL}) reads
\be
d_L(x) = c~\frac{2(1+z(x))}{H_0} \sqrt \frac{1/x - [1+z(x)]}{\Omega_{\beta,0}}~.\lb{dLx}
\ee
Combining equations \eqref{zx} and \eqref{dLx}, one can obtain numerically the function
$d_L(z)$ to be compared with observational data
\be
d_L(z) = c~\frac{2(1+z)}{H_0} \sqrt \frac{ x^{-1}(z) - (1+z) }{\Omega_{\beta,0}}~.\lb{dLz}
\ee

\begin{center}
b) redshift drift
\end{center}
The SNIa data have a large degeneracy in the $(w,\Omega_0)$ plane which can be broken using
cluster data. In our case however, due to the non standard behaviour of ``matter'', we prefer
to use other data. In view of Fig. \ref{fig1} which shows maximal deviation around
redshifts $z\sim 4$, it is interesting to use probes in this redshift range. Such probes do
not exist at the present time, though they are expected in the future (see e.g.
\cite{WMEHR12}). Here, we choose to use the redshift drift and the corresponding expected data.
The idea of redshift drift test is to collect data from the two light cones separated by
10-20 years to look for the change in redshift of a source as a function of time and it was
first noticed by Sandage \cite{sandage} and later explored by Loeb \cite{loeb}.

Contemporary technique will allow to detect this tiny effect using planned telescopes such as
the European Extremely Large Telescope (EELT) \cite{balbi,E-ELT}, the Thirty Meter Telescope
(TMT), the Giant Magellan Telescope (GMT) or even gravitational wave interferometers DECIGO/BBO
(DECi-hertz Interferometer Gravitational Wave Observatory/Big Bang Observer) \cite{DECIGO}.
Theoretically, the effect has already been investigated for the matter-dominated model
(CDM) \cite{Quercellini12}, for the $\Lambda$CDM model, for the
Dvali-Gabadadze-Porrati (DGP) brane model, for LTB models \cite{yoo}, for backreaction timescape cosmology
\cite{wiltshire}, for the axially symmetric Szekeres models \cite{marieN12}, for the Stephani
models in Paper 1 \cite{PRD13}, and for some specific dark energy models (see e.g. \cite{MP11}).

For our Stephani model II, in which we are at the center of symmetry, the redshift drift is
given by (see (\ref{redshiftdrift4}) from Appendix A with $r_0=0$)
\be
\label{redshiftdrift3}
\frac{\delta z}{\delta t} = - H_0 \left( \frac{H}{H_0} - (1+z) \right)~,\\
\ee
where \eqref{Hfull} should be used in order to express $H/H_0$.

We emphasize again that, as it was also assumed when deriving the expression for luminosity
distances, the emitting sources are assumed to be comoving.

\begin{center}
c) baryon acoustic oscillations
\end{center}

Baryon acoustic oscillations (BAO) provide us with a standard ruler \cite{AP}. Baryon oscillations generated at the
time when baryons were tightly coupled to photons and are found after decoupling in the matter power spectrum.
This gives a constraint on the universe evolution. At the present time, BAO are measured at relatively small
redshift. The constraint can be optimised with the quantity known as volume distance
\bea
D_V(z) &=& \left[ (1+z)^2 d_A^2(z) \frac{cz}{H(z)} \right]^{\frac{1}{3}}\\
&=& \left[ a_0^2 r^2(z) \frac{cz}{H(z)} \right]^{\frac{1}{3}}~.\lb{DV}
\eea
where we have used \eqref{dA} and \eqref{dL} in order to arrive at the last equality.
Using \eqref{zx}, this gives
\be
D_V(z) = \frac{c}{H_0} \left[ \frac{4}{\Omega_{\beta,0}}~z ~\frac{x^{-1}(z)-(1+z)}{h(x(z))}
                                                                         \right]^{\frac{1}{3}}~,\lb{DVz}
\ee
where $h(x)\equiv H(x)/H_0$.
The quantity $D_V$ is measured for $z=0.106, 0.2, 0.35, 0.44, 0.6, 0.73$ by the experiments
SDSS DR7 \cite{SDSSDR7}, WiggleZ \cite{WiggleZ}, and 6dF GS \cite{6dFGS}. The measurement
given by SDSS-3BOSS \cite{anderson} is $D_A(0.57) = 1408\pm 45$ Mpc and
$H(0.57) = 92.9 \pm 7.8$ km/s/Mpc.

\begin{center}
d) shift parameter
\end{center}

The location of the Cosmic Microwave Background (CMB) acoustic peaks depends on the physics
beteen us and the last scattering surface, so it provides a probe of dark energy models.
One quantity that can be used here is the so-called shift parameter \cite{bond97,Nesseris:2006er}.
It is defined as
\bea
{\cal R} &=& \sqrt{\Omega_{m,0}}~H_0~c^{-1}~(1+z_d)~d_A(z_d)\\
&=& \sqrt{\Omega_{m,0}} ~a_0 ~H_0~c^{-1}~r_d~, \lb{shift}
\eea
where $r_d\equiv r(x(z_d))$ is the coordinate distance at decoupling and we
have used \eqref{dA} and \eqref{dL} in order to arrive at the second equality.
Again, using \eqref{zx}, we obtain
\be
{\cal R} = 2 \frac{\Omega_0}{\Omega_{\beta,0}} ( x^{-1}_d - (1+z_d))^{\frac{1}{2}}~.\lb{Rx}
\ee
>From 7-year WMAP observations, the shift parameter is approximated as \cite{WMAP7}
\be
{\cal R} = 1.725\pm 0.018
\ee
Though this quantity is very accurately measured, it allows for large degeneracies which is
broken here using SNIa and redfshift drift constraints.

\subsection{Numerical results}
\setcounter{equation}{0}
\label{data}

We used a Bayesian framework to confront our Stephani  model with the cosmological observations
discussed in the previous sections. For each cosmological probe we took the likelihood function
to be Gaussian in the form
\be
p({\rm data} | \Theta) \propto \exp ( - \frac{1}{2} \chi^2),
\ee
where $\Theta$ denotes the parameters of the Stephani model and ``data" denotes generically the observed data for one of the three cosmological probes.  For the SNIa data $\chi^2$ takes the form
\bea \nonumber
\chi^2_{\rm SN}=\sum^{N}_{i,j=1} \left( C^{-1} \right)_{ij} \left(\mu_{\rm obs}(z_i)-\mu_{\rm pred}(z_i) \right) \\
\times \left(\mu_{\rm obs}(z_j)-\mu_{\rm pred}(z_j) \right)~,
\eea
where $C$ is the covariance matrix, while $\mu_{\rm obs}(z_i)$ and $\mu_{\rm pred}(z_i)$ are respectively the observed and the predicted distance modulus  of the $i^{th}$ Union2.1 SNIa \cite{Union2}. For the CMB shift parameter $\chi^2$ takes the form
\be
\chi^2_{\cal R}=\frac{({\cal R}-1.725)^2}{0.018^2}~.
\ee
Using the data for BAO at $z=0.2$ and $0.35$ taken from \cite{Percival2010}, the $\chi^2$ is given by:
\be
\chi^2_{\rm BAO}=(v_i-v^{\rm BAO}_i)(C^{-1})^{\rm BAO}_{ij}(v_j-v^{\rm BAO}_j)
\ee
where
\bea
v=\left\{\frac{r_s(z_{\rm{drag}},\Omega_m,\Omega_b; \Theta)}{D_V(0.2,\Omega_m; \Theta)},\frac{r_s(z_{\rm{drag}},\Omega_m,\Omega_b; \Theta)}{D_V(0.35,\Omega_m; \Theta)}\right\}~~
\eea
\bea v^{\rm BAO}=\left(0.1905,0.1097\right)\eea
and
\begin{eqnarray}
C^{-1}=\left(\begin{array}{cc}
30124  &  -17227\\
-17227  &  86977 \\
\end{array} \right),
\end{eqnarray}
is the inverse of the covariance matrix. In the formula above we have also used the formula for  the size of the comoving sound horizon at the baryon dragging epoch $r_s$ proposed in \cite{eisenstein}:
\be
r_{s}(z_{\rm{drag}})=153.5 \left(\frac{\Omega_b h^2}{0.02273}\right)^{-0.134}\left(\frac{ \Omega_m h^2}{0.1326}\right)^{-0.255},
\ee
with the parameters $\Omega_b h^2$ and $\Omega_m h^2$ being the physical baryon and dark matter density of the $\Lambda$CDM model, respectively.

\begin{figure}

\resizebox{80mm}{!}{\includegraphics{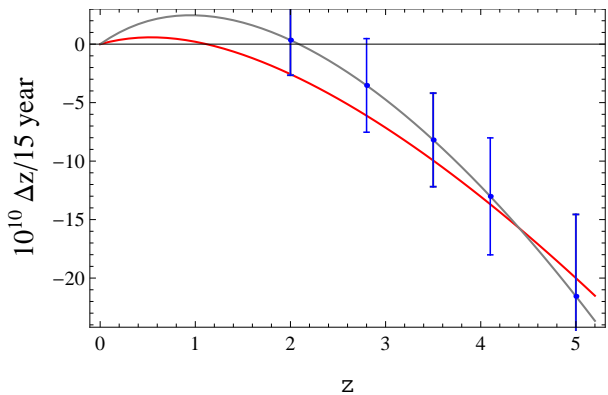}}

\caption{The redshift drift curve for the $\Lambda$CDM model (grey curve) and the Stephani
model with $\Omega_{\beta,0}=0.68$ and $w\equiv w_e(r=0)=-0.08$ found from the intersection
of the confidence intervals for SNIa, redshift drift and  BAO (red curve) (see Fig.\ref{fig3}).
Also shown are the simulated redshift drift data with their error bars from \cite{Quartin}.}
\label{fig2}
\end{figure}

For the redshift drift, we use the simulated data set presented in \cite{Quartin} (see the
blue error bars in Fig.\ref{fig2}). This data set is assumed to be centered on the
$\Lambda$CDM redshift drift curve with normally distributed errors.

With this simulated data set $\chi^2$ takes the form:
\be
\chi^2_{\rm RD}=\sum^{5}_{i=1}\frac{(\Delta z_{\rm obs}(z_i)-\Delta z_{\rm theo}(z_i))^2}{\sigma^2_i}~,
\ee
where $\Delta z_{\rm obs}(z_i)$ and $\Delta z_{\rm theo}(z_i)$ are the ``observed'' and the
predicted value of the  drift at redshift $z_i$ and $\sigma_i$ is the estimated error of the ``observed'' value of the redshift drift at  $z_i$, respectively.

In Fig.\ref{fig2} we present confidence intervals (three contours,
denoting roughly 68\%, 95\% and 99\%  confidence regions) for each observable.

\begin{figure}
\resizebox{80mm}{!}{\includegraphics{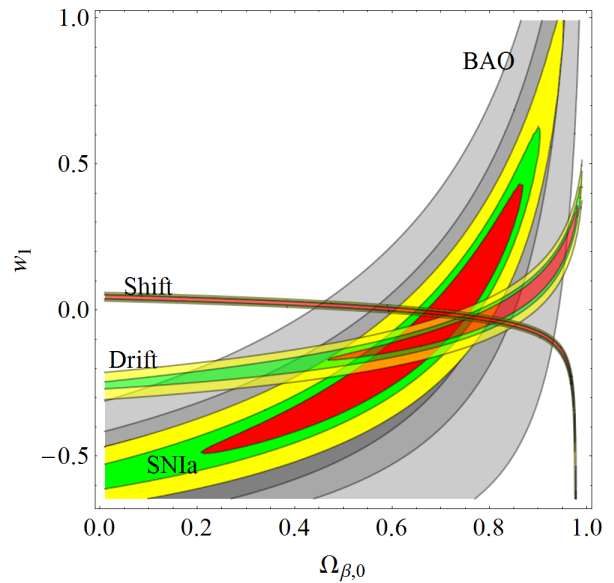}}
\caption{Confidence intervals for the SNIa, redshift drift, BAO and shift parameter.
The contours denote roughly 68\%, 95\% and 99\%  credible regions. Here a constant
equation of state parameter $w\equiv w_e(r=0)\equiv w_1$ is assumed at the center of
symmetry $r=0$.
The confidence regions from low redshifts data do not overlap with the confidence region of
the shift parameter. An additional problem is that the departure from a standard dust
behaviour $w=0$ is significant.}
\label{fig3}
\end{figure}

It is evident from  Fig. \ref{fig3} that our Stephani model fits well the data for the SNIa,
redshift drift and BAO since the related contours overlap with each other with their 1$\sigma$
CL regions. However the departure of dust from a standard behavior would render this
model unviable. In addition, it cannot comply at the same time to the CMB shift constraint.
An interesting way of overcoming the latter problem is to replace the constant barotropic
index (EoS parameter) $w$ with a function $w(a)$. Because we do not
want to change the contours obtained for SNIa, BAO, and the redshift drift, we take $w(a)$ constant
on the redshift interval from today up to $z=5$. Further, at some value of
the redshift between $z=5$ and the decoupling $z_{dec}$, we assume that the function $w(a)$
suddenly changes its value and then remains constant up to $z_{dec}$.

\begin{figure}
\resizebox{80mm}{!}{\includegraphics{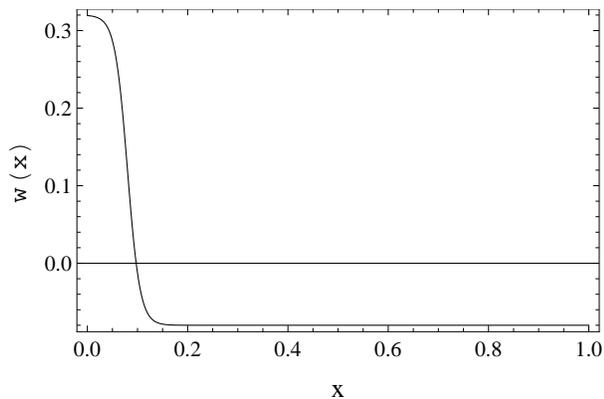}}
\caption{The time dependent equation of state parameter $w(r=0)=w(a)$
(see eq.\eqref{function}) plotted for the particular set of parameters:
$\lambda=40$, $a_{tr}=0.08$, $w_2=0.4$, $w_1=-0.08$ and $\Omega_{\beta,0}=0.68$.
Here, the transition occurs for $a\sim 0.08$ which corresponds to the redshift
$z\sim 10.49$. }
\label{fig4}
\end{figure}

An example of such function $w(a)$ fulfilling the above requirements is the
following (see Fig.\ref{fig4}):
\be
\label{function}
w(a)=w_1 +\frac{w_2}{2}\left[1+\tanh[\lambda(a_{tr}-a)]\right]~.
\ee
where $w_1$, $w_2$, $\lambda$, and $a_{tr}$ are constants.

\begin{figure}
\resizebox{80mm}{!}{\includegraphics{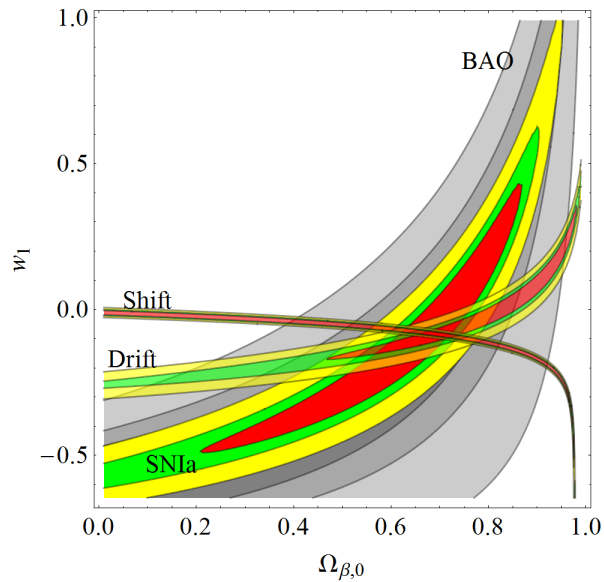}}
\caption{Confidence intervals for the SNIa, redshift drift, BAO and shift parameter for the
Stephani model with the scale factor dependent barotropic index $w(r=0)=w(a)$,
eq.\eqref{function},
where the parameters $\lambda=40$, $a_{tr}=0.08$,$w_2=0.4$ are fixed.
Again the contours denote roughly 68\%, 95\% and 99\%  confidence regions. }
\label{fig5}
\end{figure}

\begin{figure}
\resizebox{80mm}{!}{\includegraphics{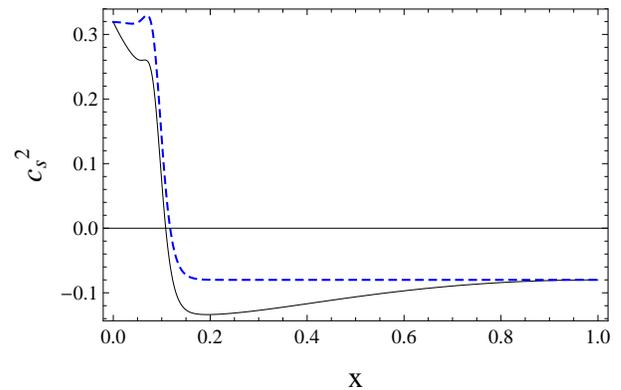}}
\caption{The speeds of sound $c_S^2(r=0,a)$ (blue dashed curve) and
$c_S^2(r,a)$ (black curve) are shown which corresponds to the barotropic index
$w(r=0)=w(a)$, eq.\eqref{function} shown on Figure \ref{fig4}. While
$c_S^2(r=0,a)$ differs from $w(a)$ in the region where $w(a)$ changes
rapidly, $c_S^2(r,a)$ includes also the effect of the pressure gradient
away from the origin.}
\label{fig6}
\end{figure}

As expected, the Stephani model with the barotropic index (\ref{function}) with $\lambda=40$, $a_{tr}=0.08$ and $w_1=-0.08, w_2=0.4$ agrees with
SNIa and BAO data and the shift parameter, while it recovers essentially the redshift
drift in a $\Lambda$CDM model (see Figs.\ref{fig2}, \ref{fig5}). Of course, this represents a
significant departure from the standard dust behaviour ($w_1 = 0$).

\section{Results and conclusions}
\setcounter{equation}{0}
\label{conclusion}

In this paper we have discussed the Stephani models of pressure-gradient spherical shells
which are complementary to the varying energy density spherical shells of the
Lema\^itre-Tolman-Bondi models. In our Stephani models there is also a spherical symmetry and
in the simplest version considered here we are assumed to be at the center of symmetry $r=0$.
A crucial difference between both spherically symmetric inhomogeneous models is the dependence
on the radial coordinate $r$ of $g_{00}$ in Stephani models. Comoving observers are no longer
following geodesics, and this is basically why a comoving perfect fluid requires a radially
dependent pressure in order to counteract the movement in the radial direction. As we have
seen, this implies that the real (``effective'') physical pressure depends on both time $t$ or
the scale factor $a$ as well as on the radial coordinate $r$.
As another general property of these models, we have seen that even if the EoS parameter
(barotropic index) $w$ (defined at $r=0$) is constant, the (adiabatic) speed of
sound will depend on both $r$ and $a$. In particular, dust ($w=0$) would acquire a negative
speed of sound and a negative equation of state parameter $w_e$ away from the origin in an
accelerating universe. The relative change of the redshift as a function of the
(generalized) scale factor $a$ will have a similar behaviour.
While we have shown that all these effects remain relatively small, though non negligible at
redshifts $z\sim 4$, on our past lightcone in a universe mimicking the cosmic history of our
universe, it is nevertheless an interesting physical property.

It is worth emphasizing the difference in the sets of models we have studied and those
studied in Ref.\cite{chris}. Our models presented in Section \ref{modelII} and dubbed Models
II in Ref.\cite{dabrowski95} further specified in Refs.\cite{stelmach01,GSS} are two-component
models with one fluid having an arbitrary barotropic equation of state at the center of
symmetry and an inhomogeneity playing another fluid with an equation of state of a domain wall
type $p=-(2/3)\varrho$. The point is that no explicit form of the scale factor is assumed, but
there is a condition $(k/a)^{\centerdot}=0$ which restricts the form of it only slightly. On the
other hand, Models I of Ref.\cite{dabrowski95} used in Ref.\cite{chris} in general do not have
such a property because for these models $(k/a)^{\centerdot}= - \Delta (\dot{a}/a^2)$ and this
reduces to Models II studied in Ref.\cite{dabrowski95} if $\Delta =0$. Besides, these models
have the scale factor assumed explicitly at the expense of having a non-barotropic equation of
state which only at the center of symmetry reduces to a barotropic one, but with a specific
form as for the network of cosmic strings with an equation of state $p = -\frac{1}{3}\varrho$.
Because of a general choice of $\Delta$ in Ref.\cite{chris}, the Models I studied there seem
to allow more freedom of the parameters which presumably lead to a larger inhomogeneity
secured by generalized EGS theorem \cite{chris99} telling us about the admittance of isotropic
radiation field to every fundamental observer in the universe.

We have also seen that our best fit model requires the barotropic index $w$ to depend on the
generalized scale factor $a$ and to be in the interval $-0.08\lesssim w(a)\lesssim 0.3$,
presumably ruling out this model. Indeed, the integrated Sachs-Wolfe effect
constraints severely any deviations from the standard dust behavior. We expect that the
formation of structure would also be strongly affected.

Another interesting issue concerns compact objects formed through gravitational collapse.
While the background perfect fluid is comoving due to its pressure gradient, it is an
interesting question whether compact objects that form out of the perfect fluids perturbations
will remain essentially comoving and for how long. A detailed study of all these problems
including the growth of perturbations can probably not be addressed analytically and is
beyond the scope of this work.

Of course, this model can always yield an observationally acceptable universe if all standard
components, including some more standard dark energy component are present and additionally if
we take a parameter $\beta$ small enough. In that case $\beta$ can be negative as well as
positive. This would be the most natural (though minimal) use of such models if observations
would point out to some slight inhomogeneity of our universe with a residual spherical
symmetry around us. However, as far as we are aware, currently the similar status is given to
LTB models - they also require standard dark energy component in the form of $\Lambda-$term
and the inhomogeneity should really be small \cite{romano}.

\section{Acknowledgements}

A.B., M.P.D., and T.D. were financed by the National Science Center Grant DEC-2012/06/A/ST2/00395.

\appendix

\section{Redshift-drift formula for a general spherically symmetric Stephani model}

We remind following Ref. \cite{PRD13} that the light emitted by the source at two different times $t_e$ and $t_e+\delta t_e$ will be observed at $t_o$ and $t_o+\delta t_o$ related by
\be
\int_{t_e}^{t_o}\frac{dt}{a(t)}=\int_{t_e+\delta t_e}^{t_o+\delta t_o}\frac{dt}{a(t)}~.
\ee
For small $\delta t_e$ and $\delta t_o$ we have
\be
\label{rel}
\frac{\delta t_e}{a(t_e)}=\frac{\delta t_o}{a(t_o)}~.
\ee
Bearing in mind (\ref{redshift}), the redshift drift has a general definition \cite{sandage,loeb}
\begin{eqnarray}
\label{redshiftdrift}
\delta z=\frac{(u_a k^a)(r_e,t_e+\delta t_e)}{(u_a k^a)(r_0,t_0+\delta t_0)}-\frac{(u_a k^a)(r_e,t_e)}{(u_a k^a)(r_0,t_0)}~,
\end{eqnarray}
and this can be calculated to first order using the expansions (for higher order expansion see Ref. \cite{PLB13})
\bea
\label{taylor1}
&&(u_a k^a)_o= (u_a k^a)(r_0,t_0)+\frac{\partial \left[(u_a k^a)(r_0,t_0)\right]}{\partial t}\delta t_0, \\
\label{taylor2}
&&(u_a k^a)_e=(u_a k^a)(r_e,t_e)+\frac{\partial \left[(u_a k^a)(r_e,t_e)\right]}{\partial t}\delta t_e~.
\eea
>From (\ref{rs}) we have
\be
\label{contrdot}
\frac{\partial}{\partial t} \left( u_a k^a \right) = - \left( \frac{1}{a} \right)^{\centerdot} - \frac{1}{4} \left( \frac{k}{a} \right)^{\centerdot} r^2~.
\ee

Applying (\ref{taylor1}), (\ref{taylor2}), and (\ref{rel}) we obtain
\bea
\label{driftSS}
\frac{\delta z}{\delta t_0} &=& \frac{\left[\left( \frac{1}{a} \right)^{\centerdot} - \frac{1}{4} \left( \frac{k}{a} \right)^{\centerdot} r^2 \right]_e}{\left[1 + \frac{1}{4} kr^2 \right]_e } a(t_e) \\
&-& \frac{\left[\left( \frac{1}{a} \right)^{\centerdot} + \frac{1}{4} \left( \frac{k}{a} \right)^{\centerdot} r^2 \right]_o}{\left[1 + \frac{1}{4} kr^2 \right]_o } a(t_0) (1+z)
\eea
>From (\ref{driftSS}) for the Models I given by (\ref{ansMI}) we have
\be
\frac{\delta z}{\delta t_0} = \frac{a(t_e) \left( \gamma + \frac{1}{4} \alpha r_e^2 \right) - a(t_0)(1+z) \left( \gamma + \frac{1}{4} \alpha r_0^2 \right)}{1 + \frac{1}{4} \frac{\alpha t_0 + \sigma}{\gamma t_0 + \delta}}~~,
\ee
while for the Models II given by (\ref{metMII}) we obtain
\be
\label{redshiftdrift4}
\frac{\delta z}{\delta t_0}=-\frac{H_0}{1+\frac{1}{4}k(t_0)r_0^2}\left[\frac{H_e}{H_0} -(1+z) \right]~,
\ee
where $H_e \equiv H(t_e) = \dot{a}(t_e)/a(t_e)$.


\begin{thebibliography}{99}

\bibitem{DE}
 V. Sahni, A. A. Starobinsky, Int. J. Mod. Phys. D \textbf{9}, 373 (2000);
 T. Padmanabhan, Phys. Rep. \textbf{380}, 235 (2003);
 E. J. Copeland, M. Sami and S. Tsujikawa, Int. J. Mod. Phys. D \textbf{15}, 1753 (2006);
 V. Sahni, A. A. Starobinsky, Int. J. Mod. Phys. D \textbf{15}, 2105 (2006);
 M. Li, X.-D. Li, S. Wang, Y. Wang, Commun. Theor. Phys. \textbf{56}, 525 (2011);
 L. Amendola, S. Tsujikawa, \emph{Dark Energy: Theory and Observations}, Cambridge University Press, 2010.

 \bibitem{UCETolman} J.-P. Uzan, C. Clarkson, and G.F.R. Ellis, Phys. Rev. Lett., {\bf 100}, 191303 (2008).

\bibitem{center} R.R. Caldwell and A. Stebbins Phys. Rev. Lett., {\bf 100}, 191302 (2008)); C. Clarkson, B. Bassett, and T. H-Ch. Lu, Phys. Rev. Lett., {\bf 101}, 011301 (2008).

\bibitem{LTB} G. Lema\^itre, Ann. Soc. Sci. Brux. A{\bf 53}, 51 (1933); R.C. Tolman, Proc. Natl. Acad. Sci. - U.S.A., {\bf 20}, 169 (1934); H. Bondi, Mon. Not. R. Astr. Soc. {\bf 107}, 410 (1947).

\bibitem{stephani} H. Stephani, Commun. Math. Phys. {\bf 4}, 137 (1967); A. Krasi\'nski, Gen. Relativ. Gravit. {\bf 15}, 673 (1983).

\bibitem{dabrowski93} M.P. D\c{a}browski, J. Math. Phys. (N.Y.) {\bf 34}, 1447 (1993).

\bibitem{dabrowski95} M.P. D\c{a}browski, Astrophys. J. {\bf 447}, 43 (1995).

\bibitem{barnes73} A. Barnes, Gen. Rel. Grav. {\bf 4}, 105 (1973).

\bibitem{krasinski} A. Krasi\'nski, {\it Inhomogeneous Cosmological Models} (Cambridge University Press, Cambridge 1997).

\bibitem{BKC} K. Bolejko, A. Krasi\'nski, C. Hellaby, and M.-N. C\'el\'erier, {\it Structures in the Universe by Exact Methods - Formation, Evolution, Interactions} (Cambridge University Press, Cambridge, England, 2010).

\bibitem{dabrowski98} M.P. D\c{a}browski and M.A. Hendry, Astrophys. J. {\bf 498}, 67 (1998).

\bibitem{LTBtests} M.-N. C\'el\'erier, Astron. Astrophys. {\bf 362}, 840 (2000); K. Tomita, Prog. Theor. Phys. {\bf 106}, 929 (2001).

\bibitem{sandage} A. Sandage, Astrophys. J. {\bf 136}, 319 (1962).

\bibitem{loeb}A. Loeb, Astrophys. J. {\bf 499}, L11 (1998).

\bibitem{chrisroy} C. Clarkson and R. Maartens, Classical Quantum Gravity {\bf 27}, 124008 (2010).

\bibitem{ellis} G.F.R. Ellis, S.D. Nel, R. Maartens, W.R. Stoeger, and A.P. Whitman, Phys. Rep. {\bf 124}, 315 (1985).

\bibitem{GSS} W. God{\l }owski, J. Stelmach, and M. Szyd{\l }owski, Classical Quantum Gravity {\bf 21}, 3953 (2004).

\bibitem{chris} R.A. Barrett and C.A. Clarkson, Classical Quantum Gravity {\bf 17}, 5047 (2000).

\bibitem{visser} M. Visser, arXiv: 1502.02758.

\bibitem{chris99} C.A. Clarkson and R.A. Barrett, Classical Quantum Gravity {\bf 16}, 3781 (1999).

\bibitem{PRD13} A. Balcerzak and M.P. D\c{a}browski, Phys. Rev. D{\bf 87}, 063506 (2013).

\bibitem{off-center14} A. Balcerzak, M.P. D\c{a}browski, and T. Denkiewicz, Astrophys. J. {\bf 792}, 92 (2014).

\bibitem{sussmann} R.A. Sussmann, J. Math. Phys. (N.Y.) {\bf 28}, 1118 (1987); {\bf 29}, 945 (1988); {\bf 29}, 1177 (1988).

\bibitem{barrow04} J.D. Barrow, Classical Quantum Gravity {\bf 21}, L79 (2004).

\bibitem{PRD05} M.P. D\c{a}browski, Phys. Rev. D{\bf 71}, 103505 (2005).

\bibitem{shell} R.A. Vanderveld, E. E. Flanagan, and I. Wasserman, Phys. Rev. D{\bf 74}, 023506 (2006); A. Krasi\'nski, C. Hellaby, K. Bolejko, and M.-N. C\'el\'erier, Gen. Relativ. Gravit. {\bf 42}, 2453 (2010); K. Bolejko, M.-N. C\'el\'erier, A. Krasi\'nski, Classical Quantum Gravity {\bf 28}, 164002 (2011).

\bibitem{tipler+krolak} F. Tipler, Phys. Lett. A \textbf{64}, 8 (1977); A. Kr\'olak, Classical Quantum Gravity \textbf{3}, 267 (1988).

\bibitem{GSFS} J.D. Barrow, Class. Quantum Grav. {\bf 21}, 5619 (2004).

\bibitem{lazkoz} L. Fernandez-Jambrina and R. Lazkoz, Phys. Rev. D{\bf 70}, 121503(R) (2004).

\bibitem{AIP10} M.P. D\c{a}browski and T. Denkiewicz, AIP Conf. Proc. {\bf 1241}, 561 (2010).

\bibitem{phantom} R.R. Caldwell, Phys. Lett. B {\bf 545}, 23 (2002); M.P. D\c{a}browski, T. Stachowiak and M. Szyd{\l }owski,
Phys. Rev. D {\bf 68}, 103519 (2003); R.R. Caldwell, M. Kamionkowski, and N.N. Weinberg, Phys. Rev. Lett. {\bf 91}, 071301 (2003);
P.H. Frampton, Phys. Lett. B {\bf 555}, 139 (2003).

\bibitem{FJ} L. Fernandez-Jambrina and R. Lazkoz, Phys. Rev. D{\bf 74}, 064030 (2006).

\bibitem{stelmach01} J. Stelmach and I. Jakacka, Classical Quantum Gravity {\bf 18}, 2643 (2001).

\bibitem{walls} M. P. D\c{a}browski, Ann. Phys. (N.Y.) {\bf 248}, 199 (1996).

\bibitem{nemiroff} R.J. Nemiroff, R. Joshi, and B.R. Patla, arXiv: 1402.4522.


\bibitem{KS} J. Kristian and R.K. Sachs, Astrophys. J. {\bf 143}, 379 (1966).

\bibitem{Union2} R. Amanullah, et al., Astrophys. J. 716, 712 (2010).

\bibitem{WMEHR12} D. Weinberg, M. Mortonson, D. Eisenstein, C. Hirata, A. Riess, E. Rozo, Phys. Rep.{\bf 530}, 87 (2013).

\bibitem{balbi} A. Balbi and C. Quercellini, Mon. Not. R. Astron. Soc. {\bf 382}, 1623 (2007).

\bibitem{E-ELT} J. Liske {\it et al.}, Monthly Not. R. Astron. Soc. {\bf 386}, 1192 (2008).

\bibitem{DECIGO} K. Yagi and N. Seto, Phys. Rev. D{\bf 83}, 044011 (2011).

\bibitem{Quercellini12} C. Quercellini, L. Amendola, A. Balbi, P. Cabella, and M. Quartin, Phys. Rep. {\bf 521}, 95 (2012).

\bibitem{yoo} C.-M. Yoo, T. Kai, and K.-I. Nakao, Phys. Rev. D{\bf 83}, 043527 (2011).

\bibitem{wiltshire} D. Wiltshire, Phys. Rev. D{\bf 80}, 123512 (2009).

\bibitem{marieN12} P. Mishra, M.-N. C\'el\'erier, and T.P. Singh, Phys. Rev. D{\bf 86}, 083520 (2012).

\bibitem{MP11} B. Moraes, D. Polarski, Phys. Rev. D{\bf 84}, 104003 (2011).

\bibitem{AP} C. Alcock and B. Paczy\'nski, Nature {\bf 281}, 358 (1979).

\bibitem{SDSSDR7} W. Percival et al., Mon. Not. R Astr. Soc. {\bf 401}, 2148 (2010);

\bibitem{WiggleZ} C. Blake et al., Mon. Not. R Astr. Soc. {\bf 415}, 2892 (2011); Mon. Not. R Astr. Soc. {\bf 418}, 1707 (2011).

\bibitem{6dFGS} F. Beutler et al., Mon. Not. R Astr. Soc. {\bf 416}, 3017 (2011)~.

\bibitem{anderson} L. Anderson et al., arXiv: 1303.4666.

\bibitem{Nesseris:2006er} S.~Nesseris and L.~Perivolaropoulos, Journ. Cosm. Astrop. Phys. {\bf 0701}, 018 (2007).


\bibitem{WMAP7} E. Komatsu et al., Astrophys. J. Suppl. {\bf 192}, 18 (2011).

\bibitem{Percival2010} W.J. Percival, B.A. Reid, D.J. Eisenstein, N.A. Bahcall,
 T. Budavari {\it et al.}, Mon. Not. Roy. Astron. Soc. {\bf 401}, 2148 (2010).

\bibitem{Quartin} M. Quartin, L. Amendola, Phys. Rev.D{\bf 81}, 043522 (2010).

\bibitem{bond97} J.R. Bond, G. Efstathiou, and M. Tegmark, Mon. Not. Roy. Astron. Soc. {\bf 291}, L33 (1997).

\bibitem{eisenstein} D.J. Eisenstein {\it et al.}, Astroph. J. {\bf 633}, 560 (2005).




\bibitem{PLB13} A. Balcerzak and M.P. D\c{a}browski, Phys. Lett. B{\bf 728}, 15 (2014).

\bibitem{romano} A.E. Romano, S. Sanes, M. Sasaki, and A.A. Starobinsky, arXiv: 1311.1476.


\end{thebibliography}
\end{document}